\documentclass[acmsmall, authorversion, nonacm]{acmart}

\AtBeginDocument{%
  \providecommand\BibTeX{{%
    \normalfont B\kern-0.5em{\scshape i\kern-0.25em b}\kern-0.8em\TeX}}}

\usepackage{graphicx}
\usepackage{subcaption}
\usepackage{tikz}
\usepackage{pgfplots}

\usepackage{hyperref} 

\usepackage{amsmath}
\usepackage{tabularx}
\usepackage{enumitem}
\usepackage{svg}

\newcommand{\name}{NGC Copilot~}

\begin{document}

\title{Adapting Foundation Models for Information Synthesis of Wireless Communication Specifications}

\author{Manikanta Kotaru}
\affiliation{%
  \institution{Microsoft}
  \country{USA}
  \address{mkotaru@microsoft.com}
}

\begin{abstract}
Existing approaches to understanding, developing and researching modern wireless communication technologies involves time-intensive and arduous process of sifting through numerous webpages and technical specification documents, gathering the required information and synthesizing it. This paper presents NextGen Communications Copilot, a conversational artificial intelligence tool for information synthesis of wireless communication specifications. The system builds on top of recent advancements in foundation models and consists of three key additional components: a domain-specific database, a context extractor, and a feedback mechanism. The system appends user queries with concise and query-dependent contextual information extracted from a database of wireless technical specifications and incorporates tools for expert feedback and data contributions. On evaluation using a benchmark dataset of queries and reference responses created by subject matter experts, the system demonstrated more relevant and accurate answers with an average BLEU score and BERTScore F1-measure of 0.37 and 0.79 respectively compared to the corresponding values of 0.07 and 0.59 achieved by state-of-the-art tools like ChatGPT. 
\end{abstract}



\maketitle

\section{Introduction}
The complexity of technical specifications presents a formidable challenge in understanding, developing, researching, or modifying modern wireless communication technologies. Researchers, practitioners, engineers and students can find themselves grappling with a multitude of acronyms and intricate terminology with information spread across a large number of documents, which can prove to be an onerous and time-consuming task to work with and develop standards-compliant systems. For example, an engineering team working on implementing registration request procedure as a part of building 5G virtual core would need to identify all the relevant technical specifications from among thousands of such documents, and understand the call flow and message formats as described in those specifications. Table~\ref{tab:story} provides several examples of such user stories.

\begin{figure}
  \centering
  
  \begin{subfigure}[b]{0.49\linewidth}
    \centering
    \includegraphics[height=3.5cm]{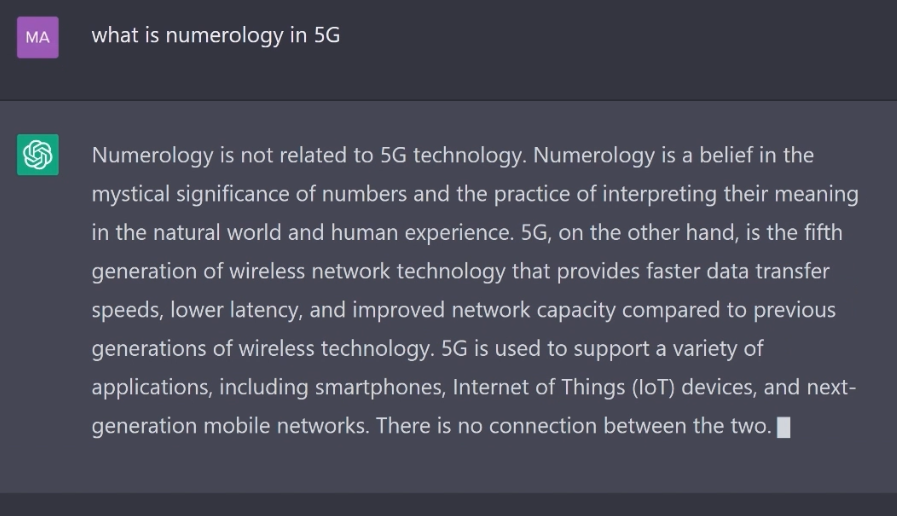}
    \caption{ChatGPT}
    \label{fig:subfigure1}
  \end{subfigure}
  \hfill
  \begin{subfigure}[b]{0.49\linewidth}
    \centering
    \includegraphics[height=3.5cm]{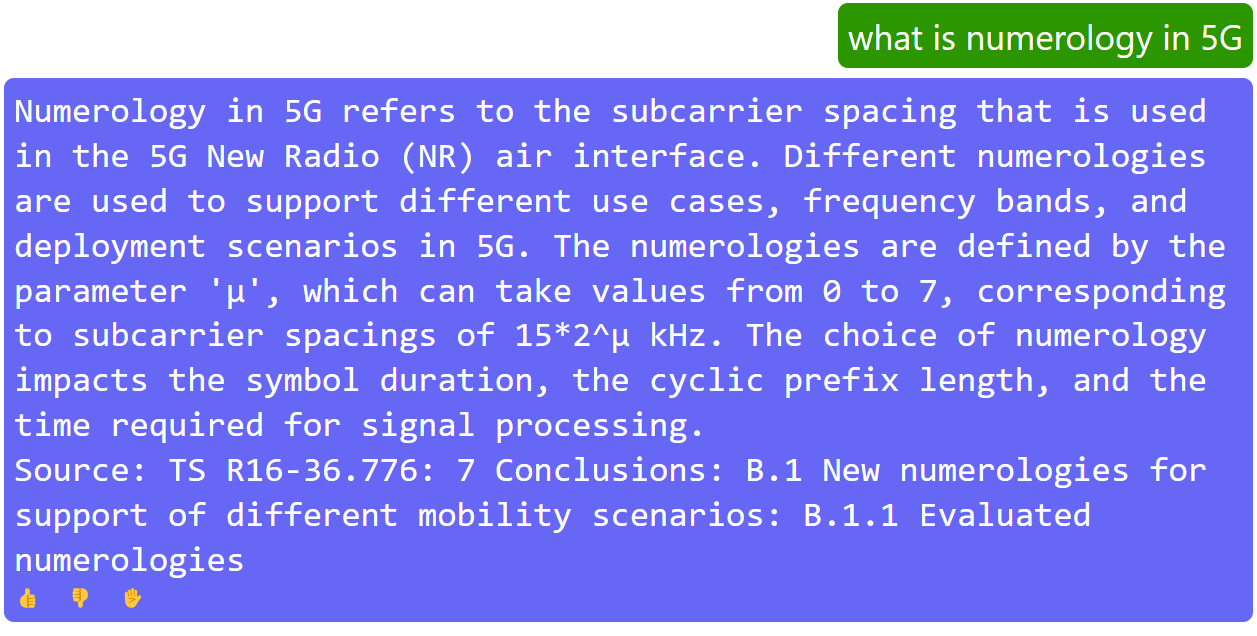}
    \caption{NextGen Communications Copilot}
    \label{fig:subfigure2}
  \end{subfigure}
  
  \caption{Comparison of responses between ChatGPT and NextGen Communications Copilot for a sample 5G-related user query}
  \label{fig:teaser}
\end{figure}

\begin{table*}
  \caption{User stories}
  \label{tab:story}
  \begin{tabularx}{\linewidth}{lX}
    \toprule
    User & Story \\
    \midrule
    As a researcher,  & I want to analyze the parameters that affect 5G MAC scheduling so that I can develop spectrally efficient resource scheduling algorithms. \\
    As an engineer, & I want to identify all technical specifications relevant to handoff and associated call flows so that I can implement handoff procedure.  \\
    As a systems integrator, & I want a clear specification of the fronthaul message format so that I can integrate open radio units with my far edge system.\\
    As an enterprise 5G user, & I want to know about slicing and its configuration parameters, so that I can efficiently use it to enhance my network services.\\
    As a student, & I want to learn about sounding reference signals so that I can understand the channel feedback mechanisms.\\  
    \bottomrule
  \end{tabularx}
\end{table*}

The current method of acquiring this information involves sifting through numerous webpages and technical specification documents. While this approach provides extensive comprehension of a topic from various sources, it can also be very time-intensive and tedious to identify multiple relevant sources, gather information from them and synthesize it~\cite{9935715}. The emergence of foundation models~\cite{foundationModels} like ChatGPT~\cite{chatgpt} presents a promising prospect for solving this problem as they represent a significant advancement in providing synthesized, readily comprehensible answers to user queries related to wireless communication specifications and technologies. 

However, despite the usefulness of state-of-the-art foundation large language models (LLMs) in answering many queries related to modern wireless communication technologies, they offer irrelevant or inaccurate responses to many of these queries.  For example, as shown in Figure~\ref{fig:teaser}(a), when prompted about ‘what is numerology in 5G’, ChatGPT (Feb 2023) describes that numerology is related to mystical significance of numbers and has no connection to 5G. Similarly, when prompted about `the number of unique values physical identity can take in 5G', it responds that `PCI consists of a 3-bit value ranging from 0 to 503', which is inaccurate and also non-sensible as 3-bit value cannot take 504 different values. 

This paper presents NextGen Communications Copilot (referred to as NGC Copilot), a conversational AI tool for information synthesis of wireless communication specifications. Similar to ChatGPT, NGC Copilot offers a question-and-answer interface, but with an enhanced ability to provide more accurate and relevant answers on topics pertaining to technical specifications of contemporary wireless communication specifications. NGC Copilot is built upon foundation LLMs and features three key additional components:

\begin{itemize}[leftmargin=*]
    \item \textbf{Domain-specific database: }Foundation models are trained on a large corpus of web data~\cite{chatgpt}. However, technical specifications and related documents for modern wireless communication technologies, eventhough publicly available, are not readily accessible~\cite{rel16} or commonly discussed within online forums, owing it to being a niche and specialized domain. This leads to missed patterns hindering the ability of state-of-the-art foundation models to generate responses with patterns relevant to wireless communication systems. To address this, the NGC Copilot system extracts and segments text from different technical specifications from 3GPP releases, WiFi standards and O-RAN~\cite{oran} to supplement the foundation models.

\item \textbf{Context Extractor: }The system utilizes a context extractor to identify the most relevant text samples from a domain-specific database and provide them as context to either a base or fine-tuned foundation model. The combined context and user query generate a response, while the system stores the specification document identifier for citation purposes. A data segmentation procedure is incorporated so that the overall prompt still fits the input size of the LLMs.

\item \textbf{Feedback: }Our system includes a feedback mechanism that allows users to seek clarification from an expert, in addition to liking or disliking the responses. When activated, the system generates an issue in a repository. The issue consists of the query, context, and response. This issue can be resolved with expert feedback and data contributions, which is  then incorporated into the domain-specific database, improving the quality of future responses.

\end{itemize}

The system is built on top of OpenAI's GPT-4 foundation model~\cite{gpt4} and 3GPP Release 16 technical specifications~\cite{rel16}. A benchmark dataset with 109 query and reference response pairs on diverse topics related to technical specifications was created. The~\name system outperformed ChatGPT, the most advanced out-of-the-box tool available for answering questions related to wireless communications specifications, with a superior average BLEU~\cite{papineni2002bleu} score of 0.37 on the benchmark dataset compared to 0.07 for ChatGPT responses for the same queries. The system also demonstrates better performance of 0.79 compared to 0.59 for ChatGPT, when evaluated using a semantic  similarity metric like BERTScore~\cite{zhang2019bertscore}. The NextGen Communication Copilot tool, benchmark dataset, and expert feedback repository will be publicly available.

\subsection{Contributions}
\begin{itemize}[leftmargin=*]
\item The paper’s main contribution lies in the application and extensive analysis of foundation models and related techniques within the context of wireless communication specifications.
\item An open-source benchmark dataset of queries and responses, created by subject matter experts, to evaluate information synthesis tools in the domain of wireless communication technologies.
\item A mechanism to enhance the system with usage through incorporation of expert feedback and data contribution, achieved by re-purposing repository issues.  
\end{itemize}
The paper also presents a few future research directions in applying foundation models in the domain of wireless systems.

\subsection{Limitations}
The system currently supports the synthesis of only textual and tabular information in the specifications but not figures and code. The approach also increases the inference cost as the user query is appended with the context information. The system is intended for use by wireless domain practitioners but not to build fully automated systems based solely on its outputs.

\section{Related Work}
 Wireless communication specifications ensure interoperability and compatibility among devices and networks, which is essential for the successful operation and adoption of wireless communication technologies~\cite{baron2018unpacking, leiponen2008competing, cooklev2004wireless}. The comprehension and development of wireless communication systems rely heavily on navigating technical specifications, whose difficulty is well-documented~\cite{veres2022study, 9935715}. Although there are tools available for managing 3GPP specifications, their functionality is restricted to accessing documents through their identification numbers~\cite{netovate}, and monitoring updates to existing or new specifications~\cite{deinleinintroducing}. Resources to share and collect information about 5G in form of articles and simulations~\cite{9935715, systemsapproach} have been developed to tackle the challenges to access, read, understand and apply wireless communication specifications. In contrast, the proposed system directly extracts pertinent information from the relevant specifications.
 
Recent approaches~\cite{holm2021bidirectional, ericssonNLP} considered natural language processing (NLP) techniques for obtaining answers to telecommunications-related questions. These approaches fine-tuned BERT~\cite{bert} language model using a small question-answering database created from 493 paragraphs, consisting of simple question-answer pairs like \{question: How high can the modulation get in LTE?, answer: 64QAM\} and \{question: What is the highest data rate in LTE in down link?, answer: 300 Mbps\}. The approach does not scale well to cover entire specification documents as it requires creating a large question-answer database. This paper, in contrast, aims at developing a conversational tool that provides synthesized information in natural human-readable form for user queries. The queries can be related to any of the technical specification documents. Further,~\name provides citation to the technical specifications and incorporates mechanism for expert feedback. This paper also contributes a benchmark dataset consisting of  detailed reference responses, rather than short categorical answers, for the purpose of evaluating information synthesis tools in wireless communication systems. To the best of our knowledge, no such dataset for wireless specifications is publicly available.

Large language models have been used in the context of network protocol specifications in tasks outside of information synthesis to uncover ambiguous or under-specified sentences in specifications and generate code for protocol~\cite{yen2021semi}, and to semantically analyze 3GPP specifications to recover protocol vulnerabilities under pre-defined threat models~\cite{9519388}. SPEC5G~\cite{karim2023spec5g} created annotated datasets for summarization and text-classification tasks by fine-tuning large language models with sentences extracted from 3GPP specifications. These approaches , based on fine-tuning alone, cannot cite to the relevant specifications or information source that provides an opportunity for the user to verify the information if needed. They lack mechanisms for expert feedback and the annotated datasets are not applicable for tasks like information synthesis to obtain relevant information for user queries. 

Large language models have become increasingly popular in recent years, with many researchers and companies exploring their potential applications. One of the earliest and most well-known examples of large language models is Google's BERT (Bidirectional Encoder Representations from Transformers), which was released in 2018~\cite{bert}. Since then, a number of  models have been developed, including Generative Pretrained Transformer-3 (GPT-3)~\cite{gpt3}, BLOOM~\cite{bloom}, XLNet~\cite{xlnet} and ChatGPT~\cite{chatgpt}. These models have been trained on broad data that can be adapted to a wide range of downstream tasks language translation, question-answering, and text summarization and have been termed as foundation models~\cite{foundationModels}. 

Foundation LLMs represent the current state-of-the-art in furnishing users with intelligible responses to their inquiries~\cite{foundationModels}. They have been adapted to numerous domain-specific tasks through fine-tuning and prompt engineering approaches~\cite{foundationModels, rietzler2019adapt, ma2019domain, diao2021taming}. Prompt engineering is widely researched to tailor LLMs to domain-specific knowledge~\cite{mialon2023augmented, arora2022ask, zhou2022docprompting, chen2023places, shao2023prompting, liu2023pre, dong2022survey, cognitiveSearch, PrivateGPT}; however, its application in the context of wireless communications remains unexplored. 

Repository hosting services like GitHub~\cite{github} have become indispensable platforms for software development and collaboration~\cite{kalliamvakou2014promises}. Traditionally, GitHub issues~\cite{githubIssues} have been used for tracking and managing bugs, tasks and feature requests. But this paper proposes to re-purpose them for obtaining and managing expert data contributions.

\section{Design}\label{sec:design}
We first provide a brief primer on foundation models and conversation AI tools based on foundation LLMs in Sec.~\ref{sec:primer}. We then describe the architecture of the system and different components in the rest of the section.

\subsection{Primer}\label{sec:primer}
\subsubsection{Large language models}
LLMs are powerful natural language processing tools that can generate fluent and coherent text. The language model is designed to predict the following word or sub-word, called tokens, by analyzing the text it has observed thus far. To utilize the machine learning models, the user's input tokens need to be transformed into a numerical representation known as "input embeddings". During the training process, the model learns how to construct these embeddings so that similar vectors represent words with similar meanings.

Majority of the state-of-the-art LLMs are based on the transformer architecture~\cite{attention}, which uses input embeddings to represent words as numbers that the model can understand. An encoder processes the input text and generates hidden states that capture its meaning and context, while a decoder generates the output sequence based on the encoded input sequence. Both input and output embeddings go through positional encoding, which helps the model understand the order of words in a sentence. During training, the model learns to generate these embeddings and guess the next word by looking at the words before it. Multiple layers of encoders and decoders are used in the transformer to improve performance on various natural language processing tasks such as language translation and conversational agents.

\subsubsection{Finetuning}
Fine-tuning involves adapting a pre-trained model to a specific domain by further training it on a new dataset. The process involves selecting a pre-trained model, preprocessing the target dataset to fit the format of the pre-trained model's input, and initializing the model with existing weights and architecture. During training, the weights of the model are updated to better fit the new data, using a chosen optimization algorithm and hyperparameters. Finetuning requires careful selection of target dataset and hyperparameters for optimal performance. This approach also required repeated training when the target dataset is updated.

\subsubsection{Prompt engineering}
Prompt engineering is a technique used in language models to fine-tune the model's output for a specific task by providing tailored prompts as inputs to the model. Prompt engineering involves crafting a specific prompt that elicits the desired response from the model. The prompt can include various elements, such as keywords, context, and formatting, and can be optimized using various techniques such as grid search or reinforcement learning. The goal is to create a prompt that provides the right amount of information to the model without being too prescriptive, allowing the model to generate accurate and relevant output.

Prompt engineering has been successfully applied in various applications like sentiment analysis, question answering and adapting a pre-trained model to new data. However, this approach requires careful consideration of the prompt's content as the model's output may be biased towards the input prompt, which can lead to inaccurate or misleading results. Further, this approach requires domain expertise to create relevant prompts to achieve good performance.

\subsection{Architecture}\label{sec:arch}
\begin{figure*}[t!]
  \centering
  \includegraphics[width=0.75\linewidth]{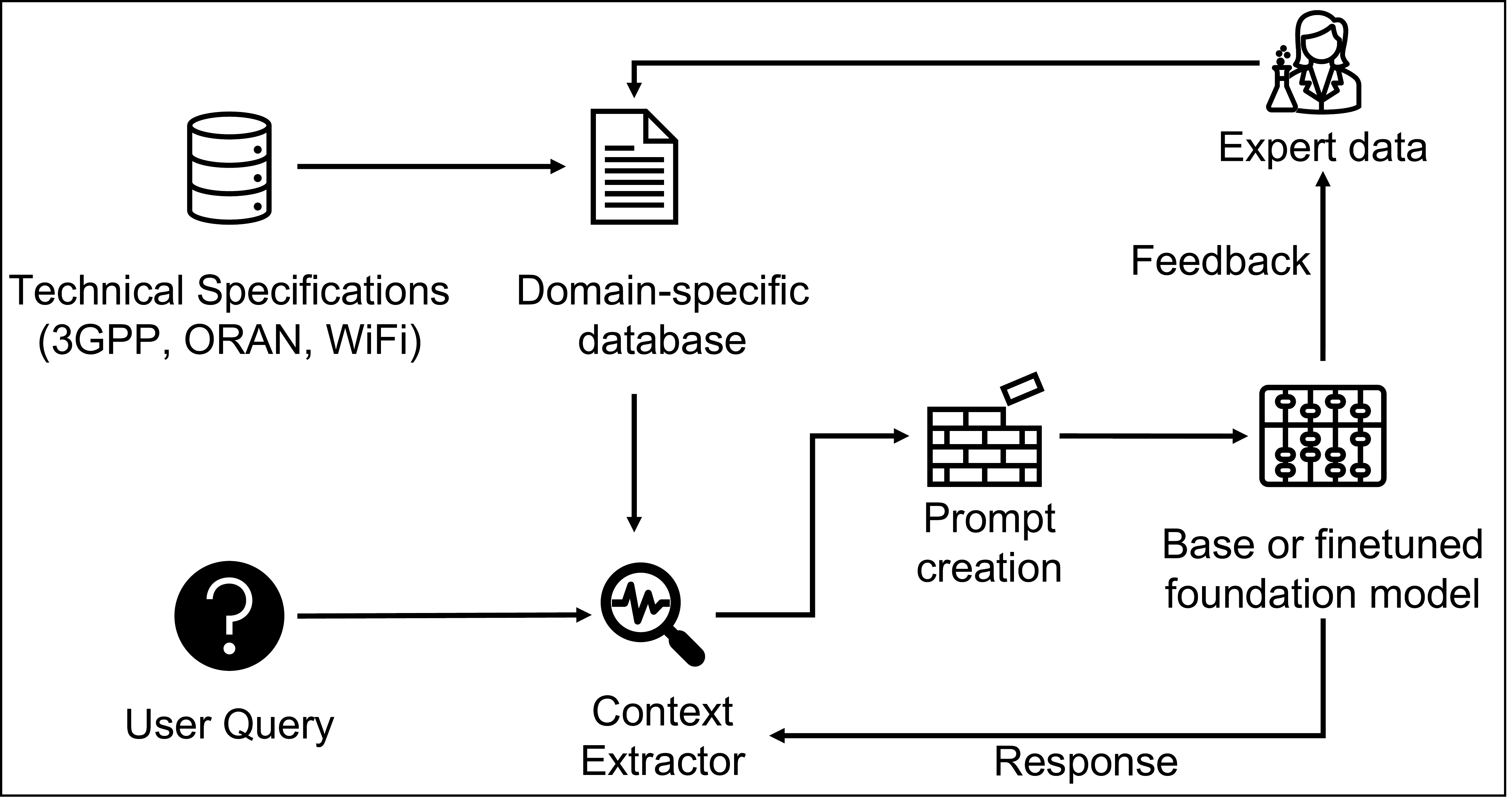}
  \caption{System architecture}
  \label{fig:arch}
\end{figure*}

Figure~\ref{fig:arch} presents the architecture of the NGC Copilot system. Prior to delving into the specifics, the following provides a concise summary of the operational procedure of the system. When a user types in a query in the message bar as shown in Figure~\ref{fig:teaser}(b), the system appends the query with relevant text samples from the domain-specific database and feeds the foundation model with the resulting combined prompt as the input. The response is displayed to the user and is used as supplemental context to subsequent queries. The response includes options for feedback as well as a button to request assistance from an expert.

\subsection{Domain-specific database}\label{sec:data}
LLMs are typically trained on a massive corpus of web data that covers a broad range of topics and domains. This training data is designed to capture the linguistic patterns and structures of natural language, enabling the models to generate coherent and contextually appropriate responses to a wide range of queries. However, domain-specific queries, like those related to wireless communication specifications, often requires relevant training data to capture the technical jargon and specialized patterns that are unique to the domain. However, such training data related to wireless  specifications is not always readily accessible. For example, 3GPP Release 16~\cite{rel16} consists of thousands of technical specifications spread across tens of compressed zip folders stored on a FTP site. Further, wireless specifications are not frequently discussed in online forums. This makes it challenging for LLMs to identify the relevant patterns and generate contextually appropriate responses. As a result, the responses generated by these models may be incomplete or irrelevant.

\subsubsection{Data collection}
The dataset used in this study was collected from 3GPP Release 16 specifications, although the technique is applicable to other 3GPP specifications, WiFi standards and O-RAN specifications. The specifications are downloaded from 3GPP FTP servers and from O-RAN specifications site.  The text from the specifications is extracted and pre-processed to remove figures, captions, code snippets, extra whitespace and certain unicode characters. 

The extracted text can optionally be used to create training samples for finetuning the base large language model. However, this paper contends that relying solely on finetuning would be inadequate for the intended application. While finetuning may facilitate content generation containing patterns pertinent to wireless communication specifications, the resultant responses may nevertheless be erroneous or incomplete. This phenomenon is acknowledged to be common even with the base LLMs~\cite{chatgpt}. Consequently, it is imperative to provide citations to the technical specifications and the data sources along with the response provided to the user. This can help ensure transparency and accountability in the use of the information synthesis tool. By citing the source of the data used to generate a response, users can better trust the information provided by the model and make informed decisions about how to use the output generated by the model. However, fine-tuning does not provide a transparent way to track the source of the data used for training. The act of citing the source of information is a crucial component that cannot be guaranteed by finetuning alone.

Prompt engineering approach presents a promising way to provide useful citations along with the response provided to the user. Prompt engineering involves designing prompts that guide the model to generate responses that are consistent with the domain-specific text by supplementing the user query with relevant data as context. By designing prompts that explicitly reference the sources of the relevant data, the model can be guided to generate responses that include useful citations. To enable such prompts, the text extracted from each of the specification documents is appended with metadata with `source' assigned to the identifier of the technical specification document. More details on the prompt design will be discussed in Sec.~\ref{sec:context_extractor}.

\subsubsection{Data segmentation}
The large language models have strict limitations on the input size creating a challenge for feeding in multiple relevant technical specification documents or even a single specification document as supplemental context to the user query. For example, text-davinci-003 model can only take an input of approximately 4097 tokens~\cite{tokenlimit}, which translates to approximately 3000 words~\cite{pricing}, a threshold that is crossed by many specification documents that run into hundreds of pages. Hence, it is necessary to segment the document into smaller chunks of reasonable size, that can be fed to the model as supplemental context.

The main challenge in segmenting domain-specific context is ensuring that the context is segmented in a way that maintains its coherence and preserves its meaning. It is important to ensure that the chunks are not arbitrary and do not introduce confusion or ambiguity. This requires careful consideration of the structure and content of the domain-specific text and an understanding of how the language model processes and integrates information.

Another challenge is ensuring that the smaller chunks are representative of the domain-specific context as a whole. In some cases, important information may be lost when the context is segmented into very small chunks. It is important to ensure that the segments are chosen in a way that preserves the salient features of the context and captures the key aspects that are relevant to the user's query.

To address these challenges, each of the documents is split along section boundaries using docx2txt library tools. Within each section, the text is recursively split into samples that are atmost $n\_words$ in length. The text within each section is not split with a hard constraint of each split sample containing exactly $n\_words$. The text within a section is split in such a way that each sample consists of multiple complete paragraphs. In other words, the sample does not end in the middle of a paragraph. Such a split respecting section, paragraph and sentence boundaries, with a reasonable length for each sample ensures that the following desired goals of chunks -- (1)  being coherent and not arbitrary, and (2) being representative of the domain-specific context as a whole -- are met. Further, a `source' string comprising the identifier of the technical specification and the section title is appended towards the end. 

\subsubsection{Word embedding}
As mentioned in Sec.~\ref{sec:primer}, the text needs to be tokenized and converted to word embedding vectors for the consumption by LLMs. The sentence-BERT all-MiniLM-L6-v2 embedding model~\cite{reimers-2019-sentence-bert} is used to transform each of the text samples into word embedding vectors. The resulting vector representation for each sample is stored in an index for efficient querying.

\subsection{Context Extractor}\label{sec:context_extractor}
The prompt engineering approach involves augmenting the user's query with relevant text from a database before feeding it to the large language model. This approach allows for the system to account for updates or modifications in the database. On the other hand, finetuning requires reasonable number (couple of hundred) of new training samples to be effective~\cite{openai-finetune}. This limits how frequently the model can be finetuned to accommodate the updates in the database. Furthermore, the prompt engineering approach also safeguards the system from generating non sequitur responses as the approach grounds the response in accordance with the relevant text from the database.

\subsubsection{Similarity search}
When the user inputs a query into~\name, the query is pre-processed, tokenized and transformed into a vector too using the same embedding model used in Sec.~\ref{sec:data}. 
The query vector is then compared to each of the word embedding vectors in the domain-specific database to identify the samples in the database that are semantically closest to the query . 

To match the user query to relevant samples in the database, a similarity metric is used. The most common similarity metric used in natural language processing (NLP) is cosine similarity. Cosine similarity measures the angle between two vectors in a high-dimensional space. If two vectors are very similar, their cosine similarity will be close to 1. If they are very different, their cosine similarity will be close to 0. Using the similarity metric, the NLP model can rank the samples in the database according to their similarity to the user query. The most similar samples are then used to create a prompt.

\subsubsection{Prompt generation}
One could directly append the most similar vectors in the database, referred to as `context', with the user query and feed the foundation LLM with the resulting prompt. However, the prompt engineering approach in combination with few-shot learning has shown promising results in several natural language processing tasks~\cite{gpt3}. Few-shot learning is enabled by feeding the foundation model with few example queries, contexts and ideal responses before providing the user query and the corresponding context. The model learns to recognize patterns in the examples and applies that knowledge to new inputs.

\subsection{Feedback}
Expert data contributions and feedback play a crucial role in improving the domain-specific responses generated by~\name due to the following reasons:
\begin{itemize}[leftmargin=*]

\item\textbf{Missing or incorrect context. }It is not always possible to include all the relevant information or context as the prompt engineering considers samples in the database that are in the top few, ranked according to the similarity metric. Expert feedback helps address this issue by providing additional context and knowledge that the model may have missed. 

\item\textbf{Queries concerning information synthesis from a large number of specifications. }User queries cannot always be answered sufficiently using small number of text samples. They may necessitate drawing upon information spanning multiple technical specifications or leveraging practical knowledge that is not explicitly documented within the specifications. Incorporating expert insights enables enriching the tool with valuable data, that enhances its ability to address these complex queries effectively.

\item \textbf{Ambiguous and under-specified protocols.}
Technical specifications are prone to contain ambiguous and under-specified sentences. These phenomenon and the resulting implications of buggy implementations and security vulnerabilities are known~\cite{yen2021semi}. Expert data would help resolve the ambiguities and provide better context.

\end{itemize}
Therefore, high-quality data contributions from experts in the domain of interest can significantly improve the performance of the model~\cite{gunasekar2023textbooks}.

Hence,~\name includes mechanisms for incorporating expert feedback and data contributions. Upon receiving a response, the user can optionally like/dislike to provide feedback on the relevance of the response. Additionally, the user can also request expert assistance by clicking a designated button, which will create a corresponding issue in a GitHub repository. This issue will contain the query, context, and response, and can be resolved through contributions from domain experts. At present, only a select few pre-identified experts are capable of resolving these issues. The expert data obtained through this process is then added to the domain-specific database and attributed to the relevant expert as its source. 

The creation of an issue in a GitHub repository ensures that the query, context, and response are stored in a central location and can be easily accessed by experts. The participation of domain experts in resolving issues ensures that the system's responses are accurate and reliable. Further, since the response and context is also posted along with the query, the expert may need to add only the information related to incomplete and incorrect portions, rather than writing an entire article answering the user query. This is in contrast to previous approaches~\cite{9935715, systemsapproach} that attempted to ease the accessibility of specification information by collaborative contributions of entire articles from experts explaining different concepts and procedures.

The attribution of responses to the expert authors ensures that experts receive recognition for their contributions, which is essential for encouraging continued participation and also creates accountability to the information that is being added to the domain-specific database. While limiting the number of experts may create a bottleneck, leading to delays in resolving the issues, the system leaves the possibility to expand the pool of experts or adopting a voting mechanism, similar to Stack Overflow~\cite{stack}, as part of future work.

\section{Evaluation}\label{sec:eval}
~\name has been built and evaluated using 3GPP Release 16 technical specifications. The text in the specifications is extracted and segmented using docx2txt~\cite{docx2txt}, antiword~\cite{antiword} and PyPDF2~\cite{pypdf} tools. The resulting database consisted of text extracted from 1663 documents, consisting of 161743 sections, and 21.8 million words. The histogram of the number of words in each section highlights a long-tail distribution. $90^{\mathrm{th}}$ percentile of the number of words per section is 360.

The $n\_words$ parameter is used to describe approximate maximum of the number of words in a text sample within domain-specific database. The choice of this parameter is guided by the observation that most of the sections have less than 360 words whereas advanced foundation models have the ability to accommodate few thousand words. We chose $n\_words$ value of 360 to segment the text to ensure that the sections are not too fragmented and so that the prompt to the foundation model can accommodate good number of filtered text samples that provide the necessary context to the user query.

The sentence-BERT all-MiniLM-L6-v2 model~\cite{reimers-2019-sentence-bert} is used for embedding the text samples and user queries into word-embedding vectors . FAISS~\cite{faiss} library is used for efficient storing of the embedding vectors in domain-specific database, and for computing cosine similarity between the query vector and all the vectors in the database. Top 3 most similar text samples are appended as supplemental context to the user query. The resulting prompt, generated using LangChain~\cite{langchain} library, is fed into the GPT-4~\cite{gpt4} foundation model. Maximum number of output tokens is set to 1000 and temperature parameter of the foundation model is set to 0 for repeatable answers to the same query. The system has been released as an app service for limited set of 18 researchers and domain experts working in the domain of cellular communication systems. 

\subsection{Benchmark dataset}\label{sec:benchmark}
To evaluate the performance of different LLM-based conversational AI tools for information synthesis of wireless communication specifications, a benchmark dataset of 109 expert queries and reference responses has been created. The responses are free form answers and not categorical. To the best of our knowledge, no such dataset for 3GPP technical specifications is publicly available. 

Each query and its associated reference response has been manually developed by subject matter experts, utilizing different sections of the 3GPP Release 16 technical specifications as the basis for their work. Multiple technical specifications, within the 1663 documents, covering different topics in the standards have been selected. Different sections have then been selected within each technical specification. In relation to any specific section, the expert has formulated one or more question and answer pairs if they ascertain that a comprehensive, independent question can be crafted, which could be answered solely based on the textual content within the section. A subset of the queries contained within the benchmark dataset drew inspiration from queries put forth by researchers utilizing the~\name. The queries span an extensive spectrum of topics that includes NR, mobility procedures, reporting formats, security, core, attributes information, authentication protocols and testing. The queries have been manually reviewed to ensure the diversity of the dataset. The dataset is used only for evaluation purposes and no training or finetuning is performed using the benchmark dataset.

The dataset will be made publicly available. Such an open source dataset potentially facilitates further research in training domain-specific language models related to wireless technologies, serving as benchmark for comparing various AI models in understanding and responding to questions about wireless technical specifications, and developing and improving conversational AI systems in the field of wireless communications. However, the authors acknowledge that the dataset, although diverse, does not cover all possible aspects of the 3GPP specifications due to its small size. Further, 3GPP standards are continually evolving while the dataset will be limited to Release 16 specifications at the time of this writing. The question answer pairs have been constrained to be self-contained to the text described within a section. However, the questions from the user could span information from multiple sections or even multiple technical specifications. Augmenting the current dataset with a larger volume of reference queries and expert data, generating questions that use information from multiple technical specifications, and exploring methodologies that could potentially facilitate the efficient expansion of the benchmark dataset by helping experts using AI tools is part of future work.

\subsection{End-to-end evaluation}
The accuracy and relevance of responses generated by the~\name is first evaluated against state-of-the-art approaches with respect to the benchmark dataset described in Sec.~\ref{sec:benchmark}. Following that, the paper performs sensitivity analysis of different components incorporated into NGC Copilot.

\subsubsection{Compared approaches: }\label{sec:compared_approaches}
~\name and the following approaches are probed with the same set of test queries selected from the benchmark dataset. 
\begin{itemize}[leftmargin=*]
    \item \textbf{ChatGPT:} 3GPP technical specifications are publicly accessible. Foundation models like GPT-4 and ChatGPT have been trained on vast corpus of publicly available web data including wireless specifications, as evidenced by relevant answers to many questions related to wireless specifications. GPT-4 and ChatGPT represent the state-of-the-art in out-of-the-box solutions available for information synthesis of wireless  specifications. For this study, the temperature and output token size of ChatGPT are set at the same values as used in the~\name.

    \item \textbf{Bing chat: }Bing chat uses Prometheus model~\cite{bing_chat} built on top of advanced GPT models. The system grounds the responses from foundation model by leveraging information extracted from web searches, ensuring a wide knowledge base that encompasses a variety of domains, including wireless technical specifications.

    \item \textbf{ChatGPT+Enterprise data: }Various closed domain question answering systems have emerged that leverage foundation models to respond to queries based on a specified collection of documents. A representative closed domain question answering system~\cite{cognitiveSearch}, which utilizes GPT-4, is considered and is pointed towards 3GPP Release 16 technical specifications.

    \item \textbf{PrivateGPT: }It is a recent popular closed domain question answering system completely relying on private models hosted locally on premises~\cite{PrivateGPT}.
\end{itemize}

\subsubsection{Metric of Merit: }%
\begin{itemize}[leftmargin=*]
    \item \textbf{Automatic evaluation: }The responses obtained from different approaches are compared against the corresponding reference responses in the benchmark dataset, employing a range of both syntactic and semantic evaluation metrics. Standard evaluation metrics like BLEU score~\cite{papineni2002bleu} and F1-measures for ROUGE-1, ROUGE-2 and ROUGE-L~\cite{rouge} metrics have been computed that effectively gauge the completeness of the response by considering the extent of shared content. 
    
    Semantic evaluation is conducted using F1-measure of BERTScore~\cite{zhang2019bertscore}, which computes the similarity usign BERT-model embeddings of the generated and reference sentences. Lastly, the cosine similarity between Sentence-BERT embeddings~\cite{reimers-2019-sentence-bert} of generated and reference responses provides another layer of semantic comparison.

\end{itemize}

\subsubsection{Results}
Table~\ref{tab:end-to-end} presents a quantitative comparison of~\name against approaches described in Sec.~\ref{sec:compared_approaches} using several standard metrics.~\name has consistently outperformed state-of-the-art approaches across all the metrics. The responses from~\name have achieved an average BLEU score of 0.37 compared to 0.07 obtained for ChatGPT responses. Similar superiority was observed in BERTScore measuring the semantic congruity of the generated responses. This can be attributed to the supplemental context provided by the context extractor. Eventhough the ChatGPT foundation model had access to wireless specifications data during its training, the smaller frequency with which such data is available in web and the ambiguity of the terms across multiple domains makes it challenging for it to learn and understand relevant text patterns.

\begin{table}[t]
\centering
\caption{End-to-end comparison using automatic evaluation metrics}
\resizebox{\textwidth}{!}{%
\begin{tabular}{|c|c|c|c|c|c|c|}
\hline
\textbf{Approach} & \textbf{BLEU} & \textbf{ROUGE-1} & \textbf{ROUGE-2} & \textbf{ROUGE-L} & \textbf{BERTScore} & Cosine \textbf{Similarity} \\
\hline
\textbf{~\name} & \textbf{0.37} & \textbf{0.60} & \textbf{0.46} & \textbf{0.59} & \textbf{0.79} & \textbf{0.72} \\
\hline
ChatGPT & 0.07 & 0.30 & 0.13 & 0.27 & 0.59 & 0.49 \\
\hline
Bing Chat & 0.07 & 0.28 & 0.13 & 0.26 & 0.58 & 0.45  \\
\hline
ChatGPT+Enterprise Data & 0.15 & 0.40 & 0.29 & 0.39 & 0.64 & 0.50  \\
\hline
PrivateGPT & 0.01 & 0.15 & 0.02 & 0.13 & 0.47 & 0.21 \\
\hline
\end{tabular}%
}
\label{tab:end-to-end}
\end{table}

Bing chat, ChatGPT+Enterprise Data and PrivateGPT approaches explicitly ground the foundation model responses with results from web-search or semantic search over the provided 3GPP technical specifications document. However, they still provide inferior responses.~\name employs segmentation procedures that are aware of the structure of wireless technical specifications documents and the distribution of words in each of the sections in those documents. Further, ~\name employs word embedding model that works well with the specifications text. Impact of the segmentation procedures and choice of word embedding models is further explored in sections ~\ref{sec:eval_segmentation} and ~\ref{sec:eval_embed_model} respectively.

To better understand how these evaluation metrics correspond to the quality of the responses, responses generated by different approaches for a sample question on `mandatory categories in QoS control rule information' are presented below. The technical specification 3GPP TS 23.203 lists the information contained in a QoS rule and which information categories are mandatory in constructing a QoS rule. The reference response is manually generated by an expert as a part of creating benchmark dataset.~\name correctly lists all the mandatory information categories while ChatGPT and Bing chat generate comprehensible but incorrect answers.  

\noindent \textbf{Question: }what are mandatory categories in QoS control rule information?

\noindent \textbf{Reference response:} The mandatory categories in QoS control rule information are: 1. Rule identifier 2. Precedence 3. Service data flow template 4. QoS class identifier Source: TS R16-23.203 Table 6.5: The QoS rule information

\noindent \textbf{NGC Copilot response:} The mandatory categories in QoS control rule information are:1. Rule identifier2. Precedence3. Service data flow template4. QoS class identifierThese categories are essential for defining and managing the Quality of Service (QoS) rules in a network.Source: TS R16-23.203 Table 6.5: The QoS rule information

\noindent \textbf{ChatGPT response:} As an AI language model, I don't have access to the latest updates. However, in general, the mandatory categories in QoS control rule information in 5G include:1. QoS Class Identifier (QCI): It defines the priority level of the traffic and the required packet delay, packet loss, and packet error rate.2. Allocation and Retention Priority (ARP): It determines the priority level of the connection and the resources allocated to it.3. Maximum Bit Rate (MBR): It specifies the maximum data rate that can be transmitted over the connection.4. Guaranteed Bit Rate (GBR): It specifies the minimum data rate that must be provided to the connection.5. Packet Delay Budget (PDB): It defines the maximum delay that can be tolerated by the connection.6. Packet Error Rate (PER): It specifies the maximum acceptable error rate for the connection.7. Priority Level: It determines the priority level of the connection in case of congestion.These categories are essential for ensuring that the network can provide the required level of service to different types of traffic and applications.

\noindent \textbf{Bing Chat response:} Quality of Service (QoS) control rule information is used to organize traffic into traffic classes or categories on the basis of whether the traffic matches specific criteria. Classifying network traffic allows you to organize traffic (that is, packets) into traffic classes or categories on the basis of whether the traffic matches specific criteria. The rules fall into three categories: user-level versus computer-level; application versus the network quintuple; and among the network quintuple. By network quintuple , we mean the source IP address, destination IP address, source port, destination port, and protocol (TCP/UDP).

\noindent \textbf{ChatGPT+Enterprise Data response:} The QoS rule information contains several pieces of information. The Category field indicates if a certain piece of information is mandatory or not for the construction of a QoS rule, i.e. if it is possible to construct a QoS rule without it. [23203-g30-120.pdf]. The mandatory categories in QoS control rule information include Rule identifier, Service data flow template, QoS class identifier, UL-maximum bitrate, DL-maximum bitrate, Guaranteed bit rate, Maximum bit rate, QoS control order, Packet filter, and IP flow mobility Routing Address [23203-g30-120.pdf]. However, the information contained in a QoS rule varies depending on the type of rule. For example, routing rule information includes Rule identifier, Routing information, Precedence, Packet filter, and IP flow mobility Routing Address [23203-g30-124.pdf]. The N6-LAN Traffic Steering Enforcement Control contains Traffic steering policy identifier(s) for steering traffic onto N6-LAN to the appropriate N6 service functions deployed by the operator [23503-gd0-91.pdf]. 

\noindent \textbf{PrivateGPT response:} The Category field indicates if a certain piece of Information is Mandatory or not, i.e., it can be constructed without the said information but there might still exist some cases where this type should never occur hence we need to have that kind of flexibility.

\subsection{Deep dive}
In this section, we conduct an ablation study by evaluating the impact of modifying different components of the~\name system. 

\subsubsection{Foundation model. }A foundation model is an essential component in~\name. Different foundation models can be used and each of these models exhibit distinct capabilities, computational costs, and complexities. We considered three foundation models in the form of GPT-4, GPT-3.5-turbo, and Text-curie-001~\cite{gpt4} models. GPT-4 is widely considered as the most advanced foundation large language model with higher computational cost for inference. GPT-3.5-turbo presents a compelling balance between performance and cost. GPT-3.5-turbo provides better performance at an order of magnitude lower cost compared to GPT-3. So, we do not consider GPT-3 model. Text-Curie-001 model presents a cheaper alternative with lower performance. 

The results from evaluating different foundation models while keeping the rest of the architecture the same are showcased in Table~\ref{tab:llm}. Integrating GPT-4 significantly outperforms other models across all the metrics, while GPT-3.5-Turbo provides an attractive alternative for cost-sensitive deployments of~\name as they provide good performance at significantly smaller cost compared to using GPT-4. There is a significant drop in the performance when the model is switched to Text-curie-001. However, it is interesting to note that even the least performing model still outperforms using ChatGPT alone in Table~\ref{tab:end-to-end} underscoring the importance of appropriate context.

\begin{table}[t]
\centering
\caption{Sensitivity analysis: Foundation models}
\resizebox{\textwidth}{!}{%
\begin{tabular}{|c|c|c|c|c|c|c|}
\hline
\textbf{Model} & \textbf{BLEU} & \textbf{ROUGE-1} & \textbf{ROUGE-2} & \textbf{ROUGE-L} & \textbf{BERTScore} & Cosine \textbf{Similarity} \\
\hline
GPT-4 & 0.37 & 0.60 & 0.46 & 0.59 & 0.79 & 0.72 \\
\hline
GPT-3.5-turbo & 0.29 & 0.55 & 0.41 & 0.54 & 0.75 & 0.69 \\
\hline
Text-curie-001 & 0.13 & 0.42 & 0.28 & 0.40 & 0.66 & 0.58 \\
\hline
\end{tabular}%
}
\label{tab:llm}
\end{table}

\subsubsection{Number of top samples to consider during similarity search}
\begin{figure}
  \centering
  \begin{subfigure}{0.48\textwidth}
    \centering
        \begin{tikzpicture}
      \begin{axis}[
        xlabel={$k$},
        ylabel={Scores},
        ymin=0,
        ymax=1,
        legend style={at={(0.98,0.02)},anchor=south east},
        grid=both,
        width=\textwidth,
        xlabel style={yshift=4pt},
        ylabel style={yshift=-10pt},
        ]
        
        \addplot[color=blue,mark=*] coordinates {
          (4, 0.383326)
          (3, 0.370580)
          (2, 0.347446)
          (1, 0.297232)
        };
        \addlegendentry{BLEU}
        
        \addplot[color=red,mark=square] coordinates {
          (4, 0.611757)
          (3, 0.603432)
          (2, 0.578835)
          (1, 0.520723)
        };
        \addlegendentry{ROUGE-1}
        
        \addplot[color=green,mark=triangle] coordinates {
          (4, 0.79)
          (3, 0.79)
          (2, 0.78)
          (1, 0.75)
        };
        \addlegendentry{BERTScore}
        
      \end{axis}
    \end{tikzpicture}
    \caption{Scores for different $k$ values.}
    \label{fig:top_k_paras}
  \end{subfigure}
  \hfill
  \begin{subfigure}{0.48\textwidth}
    \centering
      \begin{tikzpicture}[scale=0.75]
    \begin{axis}[
      ybar,
      bar width=0.35cm,
      xlabel={Model},
      ylabel={BERTScore},
      symbolic x coords={GPT-4, GPT-3.5-turbo},
      xtick=data,
      ymin=0,
      ymax=1,
      legend style={at={(0.5, 0.95)},anchor=north,legend columns=-1},
      legend cell align=left,
      grid=both,
      ylabel style={yshift=-10pt},
      ]
      
      \addplot[fill=blue] coordinates {
        (GPT-4, 0.78)
        (GPT-3.5-turbo, 0.72)
      };
      
      \addplot[fill=red] coordinates {
        (GPT-4, 0.79)
        (GPT-3.5-turbo, 0.75)
      };
      
      \legend{zero-shot, few-shot}
      
    \end{axis}
  \end{tikzpicture}
    \caption{Comparison of zero-shot and few-shot prompts.}
    \label{fig:zeroshot}
  \end{subfigure}
  \caption{Sensitivity analysis for $k$ value and few-shot learning.}
  \label{fig:combined}
\end{figure}
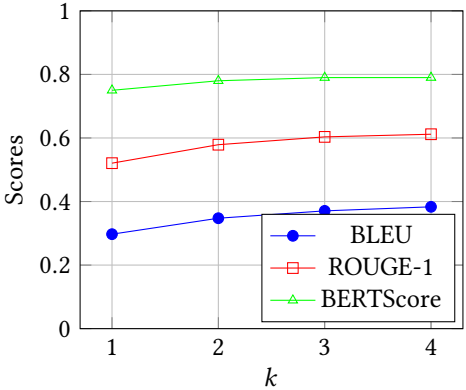
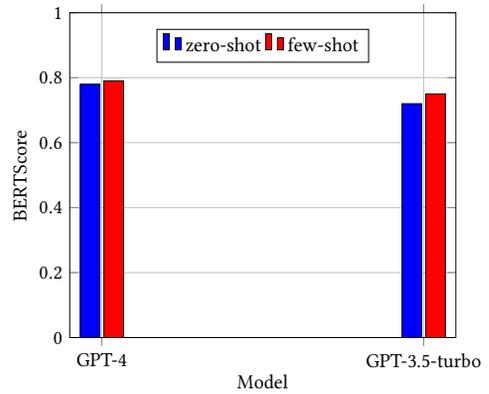

The top-\textit{k} semantically similar text samples from the domain-specific database provide the additional context to the user query. A small $k$ value results in potentially missing the text samples that are relevant to answering the user query whereas a large $k$ value results in higher inference costs due to the increased input prompt size. Figure~\ref{fig:top_k_paras} illustrates the BLEU, and F1-measures for ROUGE-1 and BERTScore metrics for the $k$ values ranging from 1 to 4. Considering the top 3 most similar text samples provides a good balance between performance and cost.

\subsubsection{Embedding model.}~\label{sec:eval_embed_model}
Embedding model plays a crucial role in~\name as it determines the context fed into  the foundation model prompt. We evaluated a variety of embedding models by integrating them into~\name while the rest of the components of the system are at their default values. These models include advanced models from Sentence-BERT~\cite{reimers-2019-sentence-bert} like all-MiniLM-L6-v2 and msmarco-distilbert-base-dot-prod-v3 as well as CPU-optimized and quantized models such as ggml-all-MiniLM-L6-v2-f16~\cite{gpt4all-embed}. All the text samples in the entire domain-specific database are embedded using each of these embedding models.~\name integrated with each of the embedding models is evaluated using the queries in the benchmark dataset. 

The BLEU score and F1-measures for the ROUGE-1 and BERTScore metrics are illustrated in Table~\ref{tab:embedding_models}. One can observe that the choice of embedding model significantly affects the performance and that the all-MiniLM-L6-v2 model outperforms other models across all the metrics. Sentence-BERT is a modification of BERT that is specifically trained for semantic similarity tasks, making it highly effective for information synthesis use case. Training a custom embedding model specifically for wireless communication specifications and evaluating its impact is an interesting research problem and is part of future work.

\begin{table}
  \centering
  \caption{Comparison of performance metrics for different embedding models.}
  \label{tab:embedding_models}
  \begin{tabular}{lccc}
    \toprule
    Embed Model & BLEU & ROUGE-1 & BERTScore \\
    \midrule
    all-MiniLM-L6-v2 & 0.37 & 0.60 & 0.79 \\
    msmarco-distilbert-base-dot-prod-v3 & 0.26 & 0.50 & 0.73 \\
   ggml-all-MiniLM-L6-v2-f16 & 0.28 & 0.52 & 0.75 \\
    \bottomrule
  \end{tabular}
\end{table}

\subsubsection{Few-shot learning. }
We conducted experiments to evaluate the performance of ~\name under both zero-shot and few-shot settings. For few-shot learning, we supplied the system with an example query, related context and ideal response, and evaluated the responses for queries from the benchmark dataset having seen these examples. For zero-shot learning, we measured how well the system could generate accurate responses to queries related to wireless communication specifications without having seen any specific examples. The results in Figure~\ref{fig:zeroshot}, show that both zero-shot and few-shot learning can be effective strategies for~\name. However, when integrated with GPT-3.5-turbo model, few-shot learning provides slightly better performance.

\subsubsection{Data segmentation approach.}~\label{sec:eval_segmentation}
Data segmentation procedures popular with other closed domain question answering systems~\cite{cognitiveSearch, PrivateGPT} where the documents are chunked into 1000 words with 100 word overlap between successive chunks of text is implemented while rest of components of the system remain the same. The performance degrades considerably with an average BLEU score and BERTScore F1-measure of 0.21 and 0.69 respectively. This demonstrates the performance improvement resulting from segmenting the text along intuitive section boundaries.

\subsubsection{Prompt selection}~\label{sec:eval_prompt}
The prompt that combines the user query and the context has significant affect on the quality of the answers for the questions in the benchmark dataset.~\name has been evaluated by keeping all the parameters including the few shot examples the same but changing the prompt alone.~\name uses the prompt `You are a helpful assistant. Use what you know already to answer the QUESTION. Improve the answer using the following pieces of CONTEXT. Always return the most relevant SOURCE.' which provides improved results compared to the prompts used by alternative approaches, as illustrated in Table~\ref{tab:prompts}.

\begin{table}
  \centering
  \caption{Comparison of performance metrics for different prompts.}
  \label{tab:prompts}
  \begin{tabular}{lccc}
    \toprule
    Prompt & BLEU & ROUGE-1 & BERTScore \\
    \midrule
    NGC Copilot prompt & 0.37 & 0.60 & 0.79 \\
    ChatGPT+Enterprise Data prompt & 0.33 & 0.56 & 0.76 \\
    PrivateGPT prompt & 0.26 & 0.51 & 0.71 \\
    \bottomrule
  \end{tabular}
\end{table}

\subsubsection{Inference cost}The average cost of each user query, for the queries in the benchmark dataset, is 6.25 cents as the context information is appended to the query. The average cost when GPT-4 model is replaced with GPT-3.5-turbo (see Table~\ref{tab:llm}), the average cost reduces to 0.35 cents without significant reduction in performance. However, the inference cost is expected to reduce in the future with the advancements in foundation LLMs and inference techniques~\cite{costInference}.

\section{Responsible AI considerations}
At the start of each session, the tool explicitly states that `we do not recommend fully automated systems based solely on the outputs and that it is important that humans are still in the loop to correct any mistakes that the system may make'. Furthermore, the answers generated by the system will be cited with specifications and there are mechanisms in place for experts to contribute data and feedback in an accountable manner. All the data used in this paper is in public domain and is not subject to copyright issues. 

However, the system may inherit any biases inherent to the text generated from the underlying foundation LLM~\cite{bommasani-etal-2020-interpreting}. For the benchmark dataset creation, guidelines were taken into consideration to be gender-neutral when possible.

\section{Future research directions}
Foundation models possess significant potential to transform the fields of wireless communication systems and mobile computing, with implications across the board from the way the systems are developed, researched, utilized and analyzed. Research in adapting foundation models to wireless communication technologies and mobile computing is still in its nascent stages. This paper aims at providing a brief overview of some of the capabilities of foundation models and present research directions in applying foundation models in these domains.

\subsection{Generation}
Foundation models have displayed remarkable ability in generating text, code and images~\cite{chatgpt, Codex, dalle} that are often difficult to distinguish from human-generated content. Adapting these capabilities to wireless technology domains to build Copilots that assist developers in generating standard-compliant code that adheres to 3GPP technical specifications, or Copilots that assist standard setting organizations in generating specifications from human intent in natural language will be transformative. Similarly, building Copilots that assist development teams in creating documentation for codebases will boost the productivity.

\subsection{Summarization and Question Answering}
Foundation models have the ability to process vast amounts of information and identify key ideas and themes of text and condensing into short meaningful summaries. Adapting foundation models to summarize a large corpus of technical papers from conferences and journals enables effective learning of a new domain. Similarly, summarization of complex specifications and codebases helps developers get a high level picture of the protocol or code functionality. It would be imperative, however, to carefully consider the copyright constraints. Foundation models' question answering capabilities allow engineers and students to iteratively mine for more and more detailed information.

\subsection{Analysis}
Foundation models perform various natural language processing tasks including text classification, sentiment analysis and named entity recognition. By training the models to recognize the key words associated with different incidents and root causes, one could build Copilots that can assist in performing root cause analysis. Similarly, adapting these models can also assist security specialists and researchers to identify security and other vulnerabilities in technical specifications by training them to recognize problematic, under-specified or ambiguous descriptions.

\subsection{Generating datasets}
To enable thriving research in different directions of applying foundation models to wireless and mobile domains, it is of utmost importance to have benchmark datasets related to different wireless specific tasks -- from information synthesis to classifying if a portion of protocol description is under-specified. It would also be important to study and develop evaluation metrics that suit these tasks specific to wireless domain.

\section{Conclusion}
The paper presents a novel application of foundation models for information synthesis of wireless communication specifications. NGC Copilot is a conversational AI tool which features a question-and-answer interface with advanced capabilities for providing accurate and relevant responses related to wireless communication specifications. The system incorporates three essential elements, including a domain-specific database, a context extractor, and a feedback mechanism. It incorporates mechanisms to overcome the limitations of foundation LLMs' input size, for citing information sources and allowing expert data contributions. 

\noindent \textbf{Acknowledgments:} The author thanks Ganesh Ananthanarayanan, Sharad Agarwal, Michael Brown, Victor Bahl, Tusher Chakraborty, Ranveer Chandra, Landon Cox, Xenofon Foukas, Pankaj Goyal, Siva Kakarla, Mustafa Kasap, Srikanth Kandula, Sanjeev Mehrotra, Munish Mehan, Sudeep Chakravartty, Bozidar Radunovic, Stefan Saroiu, Alec Wolman, Richard Whitehouse, Microsoft Networking Research and Azure for Operators Office of the CTO teams for their valuable feedback and insightful discussions.

\bibliographystyle{ACM-Reference-Format}
\bibliography{main}


\begin{thebibliography}{58}


\ifx \showCODEN    \undefined \def \showCODEN     #1{\unskip}     \fi
\ifx \showDOI      \undefined \def \showDOI       #1{#1}\fi
\ifx \showISBNx    \undefined \def \showISBNx     #1{\unskip}     \fi
\ifx \showISBNxiii \undefined \def \showISBNxiii  #1{\unskip}     \fi
\ifx \showISSN     \undefined \def \showISSN      #1{\unskip}     \fi
\ifx \showLCCN     \undefined \def \showLCCN      #1{\unskip}     \fi
\ifx \shownote     \undefined \def \shownote      #1{#1}          \fi
\ifx \showarticletitle \undefined \def \showarticletitle #1{#1}   \fi
\ifx \showURL      \undefined \def \showURL       {\relax}        \fi
\providecommand\bibfield[2]{#2}
\providecommand\bibinfo[2]{#2}
\providecommand\natexlab[1]{#1}
\providecommand\showeprint[2][]{arXiv:#2}

\bibitem[3GPP(2019)]%
        {rel16}
\bibfield{author}{\bibinfo{person}{3GPP}.} \bibinfo{year}{2019}\natexlab{}.
\newblock \bibinfo{title}{Release 16 Technical Specfications}.
\newblock
\newblock
\newblock
\shownote{https://www.3gpp.org/ftp/Specs/2022-12/Rel-16}.


\bibitem[Arora et~al\mbox{.}(2022)]%
        {arora2022ask}
\bibfield{author}{\bibinfo{person}{Simran Arora}, \bibinfo{person}{Avanika
  Narayan}, \bibinfo{person}{Mayee~F Chen}, \bibinfo{person}{Laurel~J Orr},
  \bibinfo{person}{Neel Guha}, \bibinfo{person}{Kush Bhatia},
  \bibinfo{person}{Ines Chami}, \bibinfo{person}{Frederic Sala}, {and}
  \bibinfo{person}{Christopher R{\'e}}.} \bibinfo{year}{2022}\natexlab{}.
\newblock \showarticletitle{Ask Me Anything: A simple strategy for prompting
  language models}.
\newblock \bibinfo{journal}{\emph{arXiv preprint arXiv:2210.02441}}
  (\bibinfo{year}{2022}).
\newblock


\bibitem[Azure(2023)]%
        {cognitiveSearch}
\bibfield{author}{\bibinfo{person}{Azure}.} \bibinfo{year}{2023}\natexlab{}.
\newblock \bibinfo{title}{ChatGPT+Enterprise data with Azure}.
\newblock
\newblock
\newblock
\shownote{https://github.com/Azure-Samples/azure-search-openai-demo/}.


\bibitem[Baron and Gupta(2018)]%
        {baron2018unpacking}
\bibfield{author}{\bibinfo{person}{Justus Baron} {and} \bibinfo{person}{Kirti
  Gupta}.} \bibinfo{year}{2018}\natexlab{}.
\newblock \showarticletitle{Unpacking 3GPP standards}.
\newblock \bibinfo{journal}{\emph{Journal of Economics \& Management Strategy}}
  \bibinfo{volume}{27}, \bibinfo{number}{3} (\bibinfo{year}{2018}),
  \bibinfo{pages}{433--461}.
\newblock


\bibitem[Bommasani et~al\mbox{.}(2020)]%
        {bommasani-etal-2020-interpreting}
\bibfield{author}{\bibinfo{person}{Rishi Bommasani}, \bibinfo{person}{Kelly
  Davis}, {and} \bibinfo{person}{Claire Cardie}.}
  \bibinfo{year}{2020}\natexlab{}.
\newblock \showarticletitle{{I}nterpreting {P}retrained {C}ontextualized
  {R}epresentations via {R}eductions to {S}tatic {E}mbeddings}. In
  \bibinfo{booktitle}{\emph{Proceedings of the 58th Annual Meeting of the
  Association for Computational Linguistics}}. \bibinfo{publisher}{Association
  for Computational Linguistics}, \bibinfo{address}{Online},
  \bibinfo{pages}{4758--4781}.
\newblock
\urldef\tempurl%
\url{https://doi.org/10.18653/v1/2020.acl-main.431}
\showDOI{\tempurl}
\newblock
\shownote{https://aclanthology.org/2020.acl-main.431}.


\bibitem[Bommasani et~al\mbox{.}(2021)]%
        {foundationModels}
\bibfield{author}{\bibinfo{person}{Rishi Bommasani}, \bibinfo{person}{Drew~A
  Hudson}, \bibinfo{person}{Ehsan Adeli}, \bibinfo{person}{Russ Altman},
  \bibinfo{person}{Simran Arora}, \bibinfo{person}{Sydney von Arx},
  \bibinfo{person}{Michael~S Bernstein}, \bibinfo{person}{Jeannette Bohg},
  \bibinfo{person}{Antoine Bosselut}, \bibinfo{person}{Emma Brunskill},
  {et~al\mbox{.}}} \bibinfo{year}{2021}\natexlab{}.
\newblock \showarticletitle{On the opportunities and risks of foundation
  models}.
\newblock \bibinfo{journal}{\emph{arXiv preprint arXiv:2108.07258}}
  (\bibinfo{year}{2021}).
\newblock


\bibitem[Brown et~al\mbox{.}(2020)]%
        {gpt3}
\bibfield{author}{\bibinfo{person}{Tom Brown}, \bibinfo{person}{Benjamin Mann},
  \bibinfo{person}{Nick Ryder}, \bibinfo{person}{Melanie Subbiah},
  \bibinfo{person}{Jared~D Kaplan}, \bibinfo{person}{Prafulla Dhariwal},
  \bibinfo{person}{Arvind Neelakantan}, \bibinfo{person}{Pranav Shyam},
  \bibinfo{person}{Girish Sastry}, \bibinfo{person}{Amanda Askell},
  {et~al\mbox{.}}} \bibinfo{year}{2020}\natexlab{}.
\newblock \showarticletitle{Language models are few-shot learners}.
\newblock \bibinfo{journal}{\emph{Advances in neural information processing
  systems}}  \bibinfo{volume}{33} (\bibinfo{year}{2020}),
  \bibinfo{pages}{1877--1901}.
\newblock


\bibitem[Chen et~al\mbox{.}(2023)]%
        {chen2023places}
\bibfield{author}{\bibinfo{person}{Maximillian Chen},
  \bibinfo{person}{Alexandros Papangelis}, \bibinfo{person}{Chenyang Tao},
  \bibinfo{person}{Seokhwan Kim}, \bibinfo{person}{Andy Rosenbaum},
  \bibinfo{person}{Yang Liu}, \bibinfo{person}{Zhou Yu}, {and}
  \bibinfo{person}{Dilek Hakkani-Tur}.} \bibinfo{year}{2023}\natexlab{}.
\newblock \showarticletitle{PLACES: Prompting language models for social
  conversation synthesis}.
\newblock \bibinfo{journal}{\emph{arXiv preprint arXiv:2302.03269}}
  (\bibinfo{year}{2023}).
\newblock


\bibitem[Chen et~al\mbox{.}(2021)]%
        {9519388}
\bibfield{author}{\bibinfo{person}{Yi Chen}, \bibinfo{person}{Yepeng Yao},
  \bibinfo{person}{XiaoFeng Wang}, \bibinfo{person}{Dandan Xu},
  \bibinfo{person}{Chang Yue}, \bibinfo{person}{Xiaozhong Liu},
  \bibinfo{person}{Kai Chen}, \bibinfo{person}{Haixu Tang}, {and}
  \bibinfo{person}{Baoxu Liu}.} \bibinfo{year}{2021}\natexlab{}.
\newblock \showarticletitle{Bookworm Game: Automatic Discovery of LTE
  Vulnerabilities Through Documentation Analysis}. In
  \bibinfo{booktitle}{\emph{2021 IEEE Symposium on Security and Privacy (SP)}}.
  \bibinfo{pages}{1197--1214}.
\newblock
\urldef\tempurl%
\url{https://doi.org/10.1109/SP40001.2021.00104}
\showDOI{\tempurl}


\bibitem[Cooklev(2004)]%
        {cooklev2004wireless}
\bibfield{author}{\bibinfo{person}{Todor Cooklev}.}
  \bibinfo{year}{2004}\natexlab{}.
\newblock \bibinfo{booktitle}{\emph{Wireless communication standards: A study
  of IEEE 802.11, 802.15, 802.16}}.
\newblock \bibinfo{publisher}{IEEE Standards Association}.
\newblock


\bibitem[Deinlein et~al\mbox{.}({[n.\,d.]})]%
        {deinleinintroducing}
\bibfield{author}{\bibinfo{person}{Thomas Deinlein}, \bibinfo{person}{Reinhard
  German}, {and} \bibinfo{person}{Anatoli Djanatliev}.}
  \bibinfo{year}{[n.\,d.]}\natexlab{}.
\newblock \showarticletitle{Introducing a Toolset for an easy Management of
  3GPP Specifications}.
\newblock  (\bibinfo{year}{[n.\,d.]}).
\newblock


\bibitem[Devlin et~al\mbox{.}(2018)]%
        {bert}
\bibfield{author}{\bibinfo{person}{Jacob Devlin}, \bibinfo{person}{Ming-Wei
  Chang}, \bibinfo{person}{Kenton Lee}, {and} \bibinfo{person}{Kristina
  Toutanova}.} \bibinfo{year}{2018}\natexlab{}.
\newblock \showarticletitle{Bert: Pre-training of deep bidirectional
  transformers for language understanding}.
\newblock \bibinfo{journal}{\emph{arXiv preprint arXiv:1810.04805}}
  (\bibinfo{year}{2018}).
\newblock


\bibitem[Diao et~al\mbox{.}(2021)]%
        {diao2021taming}
\bibfield{author}{\bibinfo{person}{Shizhe Diao}, \bibinfo{person}{Ruijia Xu},
  \bibinfo{person}{Hongjin Su}, \bibinfo{person}{Yilei Jiang},
  \bibinfo{person}{Yan Song}, {and} \bibinfo{person}{Tong Zhang}.}
  \bibinfo{year}{2021}\natexlab{}.
\newblock \showarticletitle{Taming pre-trained language models with n-gram
  representations for low-resource domain adaptation}. In
  \bibinfo{booktitle}{\emph{Proceedings of the 59th Annual Meeting of the
  Association for Computational Linguistics and the 11th International Joint
  Conference on Natural Language Processing (Volume 1: Long Papers)}}.
  \bibinfo{pages}{3336--3349}.
\newblock


\bibitem[Dong et~al\mbox{.}(2022)]%
        {dong2022survey}
\bibfield{author}{\bibinfo{person}{Qingxiu Dong}, \bibinfo{person}{Lei Li},
  \bibinfo{person}{Damai Dai}, \bibinfo{person}{Ce Zheng},
  \bibinfo{person}{Zhiyong Wu}, \bibinfo{person}{Baobao Chang},
  \bibinfo{person}{Xu Sun}, \bibinfo{person}{Jingjing Xu}, {and}
  \bibinfo{person}{Zhifang Sui}.} \bibinfo{year}{2022}\natexlab{}.
\newblock \showarticletitle{A Survey for In-context Learning}.
\newblock \bibinfo{journal}{\emph{arXiv preprint arXiv:2301.00234}}
  (\bibinfo{year}{2022}).
\newblock


\bibitem[Ericsson(2022)]%
        {ericssonNLP}
\bibfield{author}{\bibinfo{person}{Ericsson}.} \bibinfo{year}{2022}\natexlab{}.
\newblock \bibinfo{title}{Adopting neural language models for the telecom
  domain}.
\newblock
\newblock
\newblock
\shownote{https://www.ericsson.com/en/blog/2022/1/neural-language-models-telecom-domain}.


\bibitem[GitHub(2023a)]%
        {github}
\bibfield{author}{\bibinfo{person}{GitHub}.} \bibinfo{year}{2023}\natexlab{a}.
\newblock \bibinfo{title}{GitHub}.
\newblock
\newblock
\newblock
\shownote{https://github.com/}.


\bibitem[GitHub(2023b)]%
        {githubIssues}
\bibfield{author}{\bibinfo{person}{GitHub}.} \bibinfo{year}{2023}\natexlab{b}.
\newblock \bibinfo{title}{GitHub Issues}.
\newblock
\newblock
\newblock
\shownote{https://docs.github.com/en/issues}.


\bibitem[GPT4All(2023)]%
        {gpt4all-embed}
\bibfield{author}{\bibinfo{person}{GPT4All}.} \bibinfo{year}{2023}\natexlab{}.
\newblock \bibinfo{title}{GPT4All embeddings}.
\newblock
\newblock
\newblock
\shownote{https://docs.gpt4all.io/}.


\bibitem[Gunasekar et~al\mbox{.}(2023)]%
        {gunasekar2023textbooks}
\bibfield{author}{\bibinfo{person}{Suriya Gunasekar}, \bibinfo{person}{Yi
  Zhang}, \bibinfo{person}{Jyoti Aneja}, \bibinfo{person}{Caio
  C{\'e}sar~Teodoro Mendes}, \bibinfo{person}{Allie Del~Giorno},
  \bibinfo{person}{Sivakanth Gopi}, \bibinfo{person}{Mojan Javaheripi},
  \bibinfo{person}{Piero Kauffmann}, \bibinfo{person}{Gustavo de Rosa},
  \bibinfo{person}{Olli Saarikivi}, {et~al\mbox{.}}}
  \bibinfo{year}{2023}\natexlab{}.
\newblock \showarticletitle{Textbooks Are All You Need}.
\newblock \bibinfo{journal}{\emph{arXiv preprint arXiv:2306.11644}}
  (\bibinfo{year}{2023}).
\newblock


\bibitem[Holm(2021)]%
        {holm2021bidirectional}
\bibfield{author}{\bibinfo{person}{Henrik Holm}.}
  \bibinfo{year}{2021}\natexlab{}.
\newblock \bibinfo{title}{Bidirectional Encoder Representations from
  Transformers (BERT) for Question Answering in the Telecom Domain.: Adapting a
  BERT-like language model to the telecom domain using the ELECTRA pre-training
  approach}.
\newblock
\newblock


\bibitem[Johnson et~al\mbox{.}(2019)]%
        {faiss}
\bibfield{author}{\bibinfo{person}{Jeff Johnson}, \bibinfo{person}{Matthijs
  Douze}, {and} \bibinfo{person}{Herv{\'e} J{\'e}gou}.}
  \bibinfo{year}{2019}\natexlab{}.
\newblock \showarticletitle{Billion-scale similarity search with gpus}.
\newblock \bibinfo{journal}{\emph{IEEE Transactions on Big Data}}
  \bibinfo{volume}{7}, \bibinfo{number}{3} (\bibinfo{year}{2019}),
  \bibinfo{pages}{535--547}.
\newblock


\bibitem[Játiva et~al\mbox{.}(2022)]%
        {9935715}
\bibfield{author}{\bibinfo{person}{Grace~Montenegro Játiva},
  \bibinfo{person}{Martha Cecilia~Paredes Paredes}, {and}
  \bibinfo{person}{Julio~C. Caiza}.} \bibinfo{year}{2022}\natexlab{}.
\newblock \showarticletitle{Design of a collaborative website to study 5G
  networks}. In \bibinfo{booktitle}{\emph{2022 IEEE Sixth Ecuador Technical
  Chapters Meeting (ETCM)}}. \bibinfo{pages}{1--6}.
\newblock
\urldef\tempurl%
\url{https://doi.org/10.1109/ETCM56276.2022.9935715}
\showDOI{\tempurl}


\bibitem[Kalliamvakou et~al\mbox{.}(2014)]%
        {kalliamvakou2014promises}
\bibfield{author}{\bibinfo{person}{Eirini Kalliamvakou},
  \bibinfo{person}{Georgios Gousios}, \bibinfo{person}{Kelly Blincoe},
  \bibinfo{person}{Leif Singer}, \bibinfo{person}{Daniel~M German}, {and}
  \bibinfo{person}{Daniela Damian}.} \bibinfo{year}{2014}\natexlab{}.
\newblock \showarticletitle{The promises and perils of mining github}. In
  \bibinfo{booktitle}{\emph{Proceedings of the 11th working conference on
  mining software repositories}}. \bibinfo{pages}{92--101}.
\newblock


\bibitem[Karim et~al\mbox{.}(2023)]%
        {karim2023spec5g}
\bibfield{author}{\bibinfo{person}{Imtiaz Karim}, \bibinfo{person}{Kazi~Samin
  Mubasshir}, \bibinfo{person}{Mirza~Masfiqur Rahman}, {and}
  \bibinfo{person}{Elisa Bertino}.} \bibinfo{year}{2023}\natexlab{}.
\newblock \showarticletitle{SPEC5G: A Dataset for 5G Cellular Network Protocol
  Analysis}.
\newblock \bibinfo{journal}{\emph{arXiv preprint arXiv:2301.09201}}
  (\bibinfo{year}{2023}).
\newblock


\bibitem[LangChain(2023)]%
        {langchain}
\bibfield{author}{\bibinfo{person}{LangChain}.}
  \bibinfo{year}{2023}\natexlab{}.
\newblock \bibinfo{title}{LangChain}.
\newblock
\newblock
\newblock
\shownote{https://github.com/hwchase17/langchain}.


\bibitem[Leiponen(2008)]%
        {leiponen2008competing}
\bibfield{author}{\bibinfo{person}{Aija~Elina Leiponen}.}
  \bibinfo{year}{2008}\natexlab{}.
\newblock \showarticletitle{Competing through cooperation: The organization of
  standard setting in wireless telecommunications}.
\newblock \bibinfo{journal}{\emph{Management science}} \bibinfo{volume}{54},
  \bibinfo{number}{11} (\bibinfo{year}{2008}), \bibinfo{pages}{1904--1919}.
\newblock


\bibitem[Lin(2004)]%
        {rouge}
\bibfield{author}{\bibinfo{person}{Chin-Yew Lin}.}
  \bibinfo{year}{2004}\natexlab{}.
\newblock \showarticletitle{{ROUGE}: A Package for Automatic Evaluation of
  Summaries}. In \bibinfo{booktitle}{\emph{Text Summarization Branches Out}}.
  \bibinfo{publisher}{Association for Computational Linguistics},
  \bibinfo{address}{Barcelona, Spain}, \bibinfo{pages}{74--81}.
\newblock
\urldef\tempurl%
\url{https://aclanthology.org/W04-1013}
\showURL{%
\tempurl}


\bibitem[Linux(2023)]%
        {antiword}
\bibfield{author}{\bibinfo{person}{Linux}.} \bibinfo{year}{2023}\natexlab{}.
\newblock \bibinfo{title}{Antiword}.
\newblock
\newblock
\newblock
\shownote{https://wiki.gentoo.org/wiki/Antiword}.


\bibitem[Liu et~al\mbox{.}(2023)]%
        {liu2023pre}
\bibfield{author}{\bibinfo{person}{Pengfei Liu}, \bibinfo{person}{Weizhe Yuan},
  \bibinfo{person}{Jinlan Fu}, \bibinfo{person}{Zhengbao Jiang},
  \bibinfo{person}{Hiroaki Hayashi}, {and} \bibinfo{person}{Graham Neubig}.}
  \bibinfo{year}{2023}\natexlab{}.
\newblock \showarticletitle{Pre-train, prompt, and predict: A systematic survey
  of prompting methods in natural language processing}.
\newblock \bibinfo{journal}{\emph{Comput. Surveys}} \bibinfo{volume}{55},
  \bibinfo{number}{9} (\bibinfo{year}{2023}), \bibinfo{pages}{1--35}.
\newblock


\bibitem[Ma et~al\mbox{.}(2019)]%
        {ma2019domain}
\bibfield{author}{\bibinfo{person}{Xiaofei Ma}, \bibinfo{person}{Peng Xu},
  \bibinfo{person}{Zhiguo Wang}, \bibinfo{person}{Ramesh Nallapati}, {and}
  \bibinfo{person}{Bing Xiang}.} \bibinfo{year}{2019}\natexlab{}.
\newblock \showarticletitle{Domain adaptation with BERT-based domain
  classification and data selection}. In \bibinfo{booktitle}{\emph{Proceedings
  of the 2nd Workshop on Deep Learning Approaches for Low-Resource NLP (DeepLo
  2019)}}. \bibinfo{pages}{76--83}.
\newblock


\bibitem[Mialon et~al\mbox{.}(2023)]%
        {mialon2023augmented}
\bibfield{author}{\bibinfo{person}{Gr{\'e}goire Mialon},
  \bibinfo{person}{Roberto Dess{\`\i}}, \bibinfo{person}{Maria Lomeli},
  \bibinfo{person}{Christoforos Nalmpantis}, \bibinfo{person}{Ram Pasunuru},
  \bibinfo{person}{Roberta Raileanu}, \bibinfo{person}{Baptiste Rozi{\`e}re},
  \bibinfo{person}{Timo Schick}, \bibinfo{person}{Jane Dwivedi-Yu},
  \bibinfo{person}{Asli Celikyilmaz}, {et~al\mbox{.}}}
  \bibinfo{year}{2023}\natexlab{}.
\newblock \showarticletitle{Augmented Language Models: a Survey}.
\newblock \bibinfo{journal}{\emph{arXiv preprint arXiv:2302.07842}}
  (\bibinfo{year}{2023}).
\newblock


\bibitem[Microsoft(2023)]%
        {bing_chat}
\bibfield{author}{\bibinfo{person}{Microsoft}.}
  \bibinfo{year}{2023}\natexlab{}.
\newblock \bibinfo{title}{Bing Chat}.
\newblock
\newblock
\newblock
\shownote{https://blogs.microsoft.com/blog/2023/02/07/reinventing-search-with-a-new-ai-powered-microsoft-bing-and-edge-your-copilot-for-the-web/}.


\bibitem[Netovate(2022)]%
        {netovate}
\bibfield{author}{\bibinfo{person}{Netovate}.} \bibinfo{year}{2022}\natexlab{}.
\newblock \bibinfo{title}{3GPP tools}.
\newblock
\newblock
\newblock
\shownote{http://netovate.com/doc-search/}.


\bibitem[O-RAN(2023)]%
        {oran}
\bibfield{author}{\bibinfo{person}{O-RAN}.} \bibinfo{year}{2023}\natexlab{}.
\newblock \bibinfo{title}{Technical Specfications}.
\newblock
\newblock
\newblock
\shownote{https://www.o-ran.org/specifications}.


\bibitem[OpenAI(2022)]%
        {chatgpt}
\bibfield{author}{\bibinfo{person}{OpenAI}.} \bibinfo{year}{2022}\natexlab{}.
\newblock \bibinfo{title}{ChatGPT}.
\newblock
\newblock
\newblock
\shownote{https://openai.com/blog/chatgpt}.


\bibitem[OpenAI(2023a)]%
        {Codex}
\bibfield{author}{\bibinfo{person}{OpenAI}.} \bibinfo{year}{2023}\natexlab{a}.
\newblock \bibinfo{title}{Codex}.
\newblock
\newblock
\newblock
\shownote{https://openai.com/blog/openai-codex}.


\bibitem[OpenAI(2023b)]%
        {dalle}
\bibfield{author}{\bibinfo{person}{OpenAI}.} \bibinfo{year}{2023}\natexlab{b}.
\newblock \bibinfo{title}{Dall.E 2}.
\newblock
\newblock
\newblock
\shownote{https://openai.com/product/dall-e-2}.


\bibitem[OpenAI(2023c)]%
        {openai-finetune}
\bibfield{author}{\bibinfo{person}{OpenAI}.} \bibinfo{year}{2023}\natexlab{c}.
\newblock \bibinfo{title}{Fine-tuning}.
\newblock
\newblock
\newblock
\shownote{https://platform.openai.com/docs/guides/fine-tuning/fine-tuning}.


\bibitem[OpenAI(2023d)]%
        {gpt4}
\bibfield{author}{\bibinfo{person}{OpenAI}.} \bibinfo{year}{2023}\natexlab{d}.
\newblock \bibinfo{title}{GPT-4 Technical Report}.
\newblock
\newblock
\showeprint[arxiv]{2303.08774}~[cs.CL]


\bibitem[OpenAI(2023e)]%
        {pricing}
\bibfield{author}{\bibinfo{person}{OpenAI}.} \bibinfo{year}{2023}\natexlab{e}.
\newblock \bibinfo{title}{Pricing of the models}.
\newblock
\newblock
\newblock
\shownote{https://openai.com/pricing}.


\bibitem[OpenAI(2023f)]%
        {tokenlimit}
\bibfield{author}{\bibinfo{person}{OpenAI}.} \bibinfo{year}{2023}\natexlab{f}.
\newblock \bibinfo{title}{Token size limitations of the models}.
\newblock
\newblock
\newblock
\shownote{https://platform.openai.com/docs/models/gpt-3-5}.


\bibitem[Overflow(2023)]%
        {stack}
\bibfield{author}{\bibinfo{person}{Stack Overflow}.}
  \bibinfo{year}{2023}\natexlab{}.
\newblock \bibinfo{title}{Stack Overflow}.
\newblock
\newblock
\newblock
\shownote{https://stackoverflow.com/}.


\bibitem[Papineni et~al\mbox{.}(2002)]%
        {papineni2002bleu}
\bibfield{author}{\bibinfo{person}{Kishore Papineni}, \bibinfo{person}{Salim
  Roukos}, \bibinfo{person}{Todd Ward}, {and} \bibinfo{person}{Wei-Jing Zhu}.}
  \bibinfo{year}{2002}\natexlab{}.
\newblock \showarticletitle{Bleu: a method for automatic evaluation of machine
  translation}. In \bibinfo{booktitle}{\emph{Proceedings of the 40th annual
  meeting of the Association for Computational Linguistics}}.
  \bibinfo{pages}{311--318}.
\newblock


\bibitem[Peterson et~al\mbox{.}(2022)]%
        {systemsapproach}
\bibfield{author}{\bibinfo{person}{Larry Peterson}, \bibinfo{person}{Oguz
  Sunay}, {and} \bibinfo{person}{Bruce Davie}.}
  \bibinfo{year}{2022}\natexlab{}.
\newblock \bibinfo{title}{Private 5G: A Systems Approach}.
\newblock
\newblock
\newblock
\shownote{https://github.com/SystemsApproach/private5g}.


\bibitem[PrivateGPT(2023)]%
        {PrivateGPT}
\bibfield{author}{\bibinfo{person}{PrivateGPT}.}
  \bibinfo{year}{2023}\natexlab{}.
\newblock \bibinfo{title}{PrivateGPT}.
\newblock
\newblock
\newblock
\shownote{https://github.com/imartinez/privateGPT}.


\bibitem[PyPDF2(2023)]%
        {pypdf}
\bibfield{author}{\bibinfo{person}{PyPDF2}.} \bibinfo{year}{2023}\natexlab{}.
\newblock \bibinfo{title}{PyPDF2}.
\newblock
\newblock
\newblock
\shownote{https://pypi.org/project/PyPDF2/}.


\bibitem[Python(2023)]%
        {docx2txt}
\bibfield{author}{\bibinfo{person}{Python}.} \bibinfo{year}{2023}\natexlab{}.
\newblock \bibinfo{title}{docx2txt}.
\newblock
\newblock
\newblock
\shownote{https://pypi.org/project/docx2txt/}.


\bibitem[Reimers and Gurevych(2019)]%
        {reimers-2019-sentence-bert}
\bibfield{author}{\bibinfo{person}{Nils Reimers} {and} \bibinfo{person}{Iryna
  Gurevych}.} \bibinfo{year}{2019}\natexlab{}.
\newblock \showarticletitle{Sentence-BERT: Sentence Embeddings using Siamese
  BERT-Networks}. In \bibinfo{booktitle}{\emph{Proceedings of the 2019
  Conference on Empirical Methods in Natural Language Processing}}.
  \bibinfo{publisher}{Association for Computational Linguistics}.
\newblock
\urldef\tempurl%
\url{https://arxiv.org/abs/1908.10084}
\showURL{%
\tempurl}


\bibitem[Rietzler et~al\mbox{.}(2019)]%
        {rietzler2019adapt}
\bibfield{author}{\bibinfo{person}{Alexander Rietzler},
  \bibinfo{person}{Sebastian Stabinger}, \bibinfo{person}{Paul Opitz}, {and}
  \bibinfo{person}{Stefan Engl}.} \bibinfo{year}{2019}\natexlab{}.
\newblock \showarticletitle{Adapt or get left behind: Domain adaptation through
  bert language model finetuning for aspect-target sentiment classification}.
\newblock \bibinfo{journal}{\emph{arXiv preprint arXiv:1908.11860}}
  (\bibinfo{year}{2019}).
\newblock


\bibitem[Scao et~al\mbox{.}(2022)]%
        {bloom}
\bibfield{author}{\bibinfo{person}{Teven~Le Scao}, \bibinfo{person}{Angela
  Fan}, \bibinfo{person}{Christopher Akiki}, \bibinfo{person}{Ellie Pavlick},
  \bibinfo{person}{Suzana Ili{\'c}}, \bibinfo{person}{Daniel Hesslow},
  \bibinfo{person}{Roman Castagn{\'e}}, \bibinfo{person}{Alexandra~Sasha
  Luccioni}, \bibinfo{person}{Fran{\c{c}}ois Yvon}, \bibinfo{person}{Matthias
  Gall{\'e}}, {et~al\mbox{.}}} \bibinfo{year}{2022}\natexlab{}.
\newblock \showarticletitle{Bloom: A 176b-parameter open-access multilingual
  language model}.
\newblock \bibinfo{journal}{\emph{arXiv preprint arXiv:2211.05100}}
  (\bibinfo{year}{2022}).
\newblock


\bibitem[Shao et~al\mbox{.}(2023)]%
        {shao2023prompting}
\bibfield{author}{\bibinfo{person}{Zhenwei Shao}, \bibinfo{person}{Zhou Yu},
  \bibinfo{person}{Meng Wang}, {and} \bibinfo{person}{Jun Yu}.}
  \bibinfo{year}{2023}\natexlab{}.
\newblock \showarticletitle{Prompting Large Language Models with Answer
  Heuristics for Knowledge-based Visual Question Answering}.
\newblock \bibinfo{journal}{\emph{arXiv preprint arXiv:2303.01903}}
  (\bibinfo{year}{2023}).
\newblock


\bibitem[Sunyan(2023)]%
        {costInference}
\bibfield{author}{\bibinfo{person}{Sunyan}.} \bibinfo{year}{2023}\natexlab{}.
\newblock \bibinfo{title}{The Economics of Large Language Models}.
\newblock
\newblock
\newblock
\shownote{https://sunyan.substack.com/p/the-economics-of-large-language-models}.


\bibitem[Vaswani et~al\mbox{.}(2017)]%
        {attention}
\bibfield{author}{\bibinfo{person}{Ashish Vaswani}, \bibinfo{person}{Noam
  Shazeer}, \bibinfo{person}{Niki Parmar}, \bibinfo{person}{Jakob Uszkoreit},
  \bibinfo{person}{Llion Jones}, \bibinfo{person}{Aidan~N Gomez},
  \bibinfo{person}{{\L}ukasz Kaiser}, {and} \bibinfo{person}{Illia
  Polosukhin}.} \bibinfo{year}{2017}\natexlab{}.
\newblock \showarticletitle{Attention is all you need}.
\newblock \bibinfo{journal}{\emph{Advances in neural information processing
  systems}}  \bibinfo{volume}{30} (\bibinfo{year}{2017}).
\newblock


\bibitem[Veres(2022)]%
        {veres2022study}
\bibfield{author}{\bibinfo{person}{Flavia-Denisa Veres}.}
  \bibinfo{year}{2022}\natexlab{}.
\newblock \emph{\bibinfo{title}{A study into the usability of 3GPP technical
  specifications}}.
\newblock \bibinfo{thesistype}{Master's\ thesis}. \bibinfo{school}{University
  of Twente}.
\newblock


\bibitem[Yang et~al\mbox{.}(2019)]%
        {xlnet}
\bibfield{author}{\bibinfo{person}{Zhilin Yang}, \bibinfo{person}{Zihang Dai},
  \bibinfo{person}{Yiming Yang}, \bibinfo{person}{Jaime Carbonell},
  \bibinfo{person}{Russ~R Salakhutdinov}, {and} \bibinfo{person}{Quoc~V Le}.}
  \bibinfo{year}{2019}\natexlab{}.
\newblock \showarticletitle{Xlnet: Generalized autoregressive pretraining for
  language understanding}.
\newblock \bibinfo{journal}{\emph{Advances in neural information processing
  systems}}  \bibinfo{volume}{32} (\bibinfo{year}{2019}).
\newblock


\bibitem[Yen et~al\mbox{.}(2021)]%
        {yen2021semi}
\bibfield{author}{\bibinfo{person}{Jane Yen}, \bibinfo{person}{Tam{\'a}s
  L{\'e}vai}, \bibinfo{person}{Qinyuan Ye}, \bibinfo{person}{Xiang Ren},
  \bibinfo{person}{Ramesh Govindan}, {and} \bibinfo{person}{Barath Raghavan}.}
  \bibinfo{year}{2021}\natexlab{}.
\newblock \showarticletitle{Semi-automated protocol disambiguation and code
  generation}. In \bibinfo{booktitle}{\emph{Proceedings of the 2021 ACM SIGCOMM
  2021 Conference}}. \bibinfo{pages}{272--286}.
\newblock


\bibitem[Zhang et~al\mbox{.}(2019)]%
        {zhang2019bertscore}
\bibfield{author}{\bibinfo{person}{Tianyi Zhang}, \bibinfo{person}{Varsha
  Kishore}, \bibinfo{person}{Felix Wu}, \bibinfo{person}{Kilian~Q Weinberger},
  {and} \bibinfo{person}{Yoav Artzi}.} \bibinfo{year}{2019}\natexlab{}.
\newblock \showarticletitle{Bertscore: Evaluating text generation with bert}.
\newblock \bibinfo{journal}{\emph{arXiv preprint arXiv:1904.09675}}
  (\bibinfo{year}{2019}).
\newblock


\bibitem[Zhou et~al\mbox{.}(2022)]%
        {zhou2022docprompting}
\bibfield{author}{\bibinfo{person}{Shuyan Zhou}, \bibinfo{person}{Uri Alon},
  \bibinfo{person}{Frank~F Xu}, \bibinfo{person}{Zhiruo Wang},
  \bibinfo{person}{Zhengbao Jiang}, {and} \bibinfo{person}{Graham Neubig}.}
  \bibinfo{year}{2022}\natexlab{}.
\newblock \showarticletitle{Docprompting: Generating code by retrieving the
  docs}.
\newblock \bibinfo{journal}{\emph{arXiv preprint arXiv:2207.05987}}
  (\bibinfo{year}{2022}).
\newblock


\end{thebibliography}

\end{document}